\documentstyle[a4,12pt]{article}
\def \d{\partial}

\def \r{{\bf r}}
\def \be{\begin{equation}}
\def \ee{\end{equation}}

\def \1{{\bf 1}}
\def \aa{\alpha}
\def \bb{\beta}
\def \gg{\gamma}
\def \dd{\delta}

\def \ll{\lambda}
\def \om{\omega}
\def \ss{\sigma}
\def \bra{{\bar a}}
\def \brb{{\bar b}}

\def \bre{{\bar e}}

\def \brg{{\bar g}}
\def \brh{{\bar h}}
\def \brj{{\bar j}}
\def \brs{{\bar s}}
\def \bom{{\bar \om}}

\def \BA{{\bar A}}

\def \BC{{\bar C}}
\def \BD{{\bar D}}
\def \BR{{\bar R}}
\def \GG{\Gamma}
\def \LL{\Lambda}
\def \BGG{{\bar \GG}}
\def \BT{{\bar T}}
\def \hr{{\hat r}}
\def \Tr{\mbox{Tr}}
\def \fr{\frac}

\def \CR{{\cal R}}
\def \BCR{{\bar\CR}}

\def \hD{{\hat D}}
\newfont{\st}{cmr7}

\begin{document}


\title{ Variable Light-Cone Theory of Gravity}

\author{ I.T.~Drummond,\\
DAMTP\\ University of Cambridge\\ Silver Street\\ Cambridge, CB3~9EW, UK}
\maketitle

\begin{abstract}
We show how to reformulate Variable Speed of Light Theories (VSLT) in  
a covariant fashion as Variable Light-Cone Theories (VLCT) by introducing 
two vierbein bundles each associated with a distinct metric. The basic 
gravitational action relates to one bundle while matter propagates relative 
to the other in a conventional way. The variability of the speed of light
is represented by the variability of the matter light-cone relative to the 
gravitational light-cone. The two bundles are related locally by an element 
$M$, of $SL(4,R)$~. The dynamics of the field $M$ is that of a $SL(4,R)$-sigma 
model gauged with respect to local (orthochronous) Lorentz transformations on 
each of the bundles. Only the ``massless'' version of the model with a single
new coupling, $F$, that has the same dimensions as Newton's constant $G_N$, 
is considered in this paper. When $F$ vanishes the theory reduces to standard 
General Relativity.

We verify that the modified Bianchi identities of the model are consistent with 
the standard conservation law for the matter energy-momentum tensor in
its own background metric.

The implications of the model for some simple applications are examined.
We discuss the Newtonian limit, the appropriate
generalisations of the flat FRW universe and the spherically symmetric
static solution. We conclude that the variability of the speed of light
in the early universe is a possible homogenising mechanism. However an
examination of the post-Newtonian approximation shows that on the basis
of the results of VLBI measurements, the new coupling satisfies
$F/G_N<3.2\times 10^{-4}$~. We point out that other ``massive'' versions
of the model with different asymptotic properties may still permit
consistency with General Relativity at large distances while predicting
departures at short distances.

\end{abstract}
\vfill
DAMTP-99-102
\pagebreak

\section{Introduction}
Albrecht, Magueijo and Barrow \cite{AM}-\cite{B2} have proposed Variable
Speed of Light Theories (VSLT) to provide a stable basis for the observed
isotropy and flatness of the universe. In these models the speed of light 
is parametrised as a kind of equation of state tied to the radius of the universe.
They demonstrate that with the appropriate dependence of the velocity
of light it is possible to create a cosmic dynamics containing
long time attractors for the evolution of the universe that provide
an explanation for its current state. However the formulation of the theories is 
not obviously covariant and the implication of the theories for regimes other than 
the early universe is not immediately clear.

Moffat \cite{M1,M2} earlier presented related ideas and Clayton and Moffat \cite{CM} have proposed 
an ingenious bi-metric theory that overcomes some of the problems in formulating covariant VSLT. 
This is more fully dynamical than the bi-metric theory of Rosen \cite{R1,R2,R3}.

The possibility of ``superluminal propagation'' induced by quantum effects has been
discussed previously \cite{ITD}-\cite{GS}. The results are intriguing but remain controversial.

In a spirit of exploration  we propose in this paper a theory with variable 
speed of light that is both geometrical in structure and explicitly covariant in the same
way as standard General Relativity. Furthermore the energy momentum tensor of 
matter is conserved in the conventional way. In this way it achieves some of the same 
aims as ref \cite{CM}, which appeared while this work was in preparation.
The proposal is to introduce into the space-time manifold, two vierbein bundles. 
One is associated with matter and the other with gravity. The matter vierbein 
can be strained relative to the gravitational vierbein giving a geometrical 
meaning to variability of the speed of light. The propagation of light is of 
course associated with the light-cone of the matter vierbein. Because of this 
unconventional approach to light and matter propagation we will refer to the 
Variable Light-Cone Theory (VLCT). It has a more general structure than just a 
varying speed of light. The theory in some aspects has the flavour of the bi-metric 
theories \cite{CM,R1,R2,R3} but the emphasis is more on the relationship of the 
two vierbein bundles rather than on the two implied metrics.

The dynamics of the theory is specified by giving gravity its
standard curvature based action, $I_G$, and matter its standard
action, $I_M$, in the appropriate metric, and by introducing a linking
action, $I_L$, that controls the relationship between the two
vierbein bundles. The full action, $I$, is the sum of all three terms,
\be
I=I_G+I_L+I_M~~.
\ee

\section{General Structure}

The theory is most clearly formulated in the vierbein formalism and makes
fundamental use of the resulting Lorentz gauge invariance.
We introduce a vierbein bundle appropriate to gravity, $\{e_{\mu a}\}$,
with the associated metric
\be
g_{\mu\nu}=e_{\mu a}e_{\nu}^{~~a}~~,
\ee
where raising and lowering of $a$-indices is carried out with the standard
Lorentz metric $\eta_{ab}=\{1,-1,-1,-1\}$~. The inverse vierbein is $\{e^{a\mu}\}$
so that
\be
e_{\mu a}e^{a\nu}=\dd^{\nu}_{\mu}~~,~~~~~~~~e^{a\mu}e_{\mu b}=\dd^{a}_{b}~~,
\ee
and
\be
g^{\mu\nu}=e^{a\mu}e_a^{~~\nu}~~.
\ee
The vierbein associated with matter is $\{\bre_{\mu\bra}\}$ and the raising and lowering
of $\bra$-indices is by means of the Lorentz metric, $\eta_{\bra\brb}=\{1,-1,-1,-1\}$~.
The associated metric is
\be
brg_{\mu\nu}=\bre_{\mu\bra}\bre_{\nu}^{~~\bra}~~.
\ee
The two vierbein bundles are related by a local linear transformation
\be
\bre_{\mu\bra}=e_{\mu a}M^a_{~~\bra}~~,
\ee
where we assume that the matrix $M$ is an element of $SL(4,R)$~. This implies 
that the volume elements in the two bundles are the same. We denote the common value
of the two determinants by $J$~.
\be
J=\det\{e_\mu^{~~a}\}=\det\{\bre_\mu^{~~\bra}\}~~.
\ee
It would be interesting to know if the assumption that $\det M=1$ could be relaxed. 
It seems to play a role in ensuring the conservation of the matter energy-momentum
tensor.

We denote the inverse matrix by $M^{\bra}_{~~a}$ so that
\be
M^a_{~~\bra}M^{\bra}_{~~b}=\dd^a_b~~,~~~~~~~~M^{\bra}_{~~a}M^a_{~~\brb}=\dd^{\bra}_{\brb}~~.
\ee

We introduce vierbein connections for both bundles and associated coordinate connections.
The relationship between them is achieved by requiring that the vierbeins are covariantly
constant in the appropriate way.
\be
D_\mu e_{\nu a}=\d_\mu e_{\nu a}+\om_{\mu a}^{~~~b}e_{\nu b}-\GG^\ll_{\mu\nu}e_{\ll a}=0~~,
\ee
and
\be
\BD_\mu\bre_{\nu\bra}
=\d_\mu\bre_{\nu\bra}+\bom_{\mu\bra}^{~~~\brb}\bre_{\nu\brb}-\BGG^\ll_{\mu\nu}\bre_{\ll \bra}=0~~.
\ee
The requirement that $\eta_{a b}$ and $\eta_{\bra\brb}$ be covariantly constant 
implies that $\om_{\mu ab}=-\om_{\mu ba}$ and $\bom_{\mu\bra\brb}=-\bom_{\mu\brb\bra}$~.

It is convenient to define a covariant derivative of $M$ that includes both the right and left 
vierbein connections,
\be
D_\mu M^a_{~~\bra}=\d_\mu M^a_{~~\bra}+\om^{~a}_{\mu~~b}M^b_{~~\bra}
                     -M^a_{~~\brb}~\bom_{\mu~~\bra}^{~\brb}~~.
\ee
However in differentiating a second time the appropriate spatial connection, 
$\GG^\ll_{\mu\nu}$, must be used. The covariant deriviative $\BD_\mu$ 
can be extended in a similar way. Its effect on $M$ is the same as that of $D_\mu$ but
a second differentiation must use the spatial connection $\BGG^\ll_{\mu\nu}$~. 

Gravitational curvature tensors are defined so that
\be
\left[D^L_\mu,D^L_\nu\right]V_a=R_{ab\mu\nu}V^b~~,
\ee
and
\be
\left[D^R_\mu,D^R_\nu\right]V_\bra=\BR_{\bra\brb\mu\nu}V^\brb~~,
\ee
where $D^L_\mu$ includes the left vierbein connection field, $\om_{\mu ab}$ but not
the spatial connection, $\GG^\ll_{\mu\nu}$~. Similarly $D^R_\mu$ includes the only the
right vierbein connection, $\bom_{\mu \bra\brb}$~. We have then
\be
R_{ab\mu\nu}=\d_\mu\om_{\nu ab}-\d_\nu\om_{\mu ab}
 +\om_{\mu a}^{~~~c}~\om_{\nu cb}-\om_{\nu a}^{~~~c}~\om_{\mu cb}~~,
\label{CURV1}
\ee
with a similar definition for $\BR_{\bra\brb\mu\nu}$~. It follows that
\be
\left[D_\mu,D_\nu\right]M^a_{~~\bra}
          =R^a_{~~b\mu\nu}M^b_{~~\bra}-M^a_{~~\brb}\BR^\brb_{~~\bra\mu\nu}
       -2C^\ll_{\mu\nu}D_\ll M^a_{~~\bra}~~,
\label{CURV2}
\ee
where
\be
C^\ll_{\mu\nu}=\fr{1}{2}\left(\GG^\ll_{\mu\nu}-\GG^\ll_{\nu\mu}\right)~~.
\ee
The quantity $C^\ll_{\mu\nu}$ is the torsion tensor in the coordinate basis.

\section{Gravitational Action}

The gravitational action is the standard action
\be
I_G=-\fr{1}{16\pi G}\int d^4xJR~~,
\ee
where
\be
R=e^{a\mu}e^{b\nu}R_{ab\mu\nu}~~.
\ee
The vierbeins and the connections are treated as independent variables.
If we vary the former then
\be
\dd I_G=\fr{1}{8\pi G}\int d^4xJ\dd e_{\ss c}\left(e^{c\mu}R^\ss_{~~\mu}-\fr{1}{2}e^{c\ss}R\right)~~.
\ee
The variation of the vierbein connection yields
\be
\dd I_G=-\fr{1}{16\pi G}\int d^4xJe^{a\mu}e^{b\nu}\left(D^L_\mu\dd\om_{\nu ab}-D^L_\nu\dd\om_{\mu ab}\right)~~.
\ee
Switching to the full covariant derivative we get
\be
\dd I_G=-\fr{1}{16\pi G}\int d^4xJe^{a\mu}e^{b\nu}
 \left(D_\mu\dd\om_{\nu ab}-D_\nu\dd\om_{\mu ab}+2C^\ll_{\mu\nu}\dd\om_{\ll ab}\right)~~.
\ee
Using the result
\be
JD_\mu V^\mu=\d_\mu(JV^\mu)+2JC^\ll_{\ll\mu}V^\mu~~.
\ee
we can integrate by parts and obtain finally
\be
\dd I_G=-\fr{1}{8\pi G}\int d^4xJe^{a\mu}e^{b\nu}\left(C^\ll_{\ll\mu}\dd^\ss_\nu
    -C^\ll_{\ll\nu}\dd^\ss_\mu+C^\ss_{\mu\nu}\right)\dd\om_{\ss ab}~~.
\ee
If there is no other interaction in the theory we can deduce
from the vanishing of these variations that $C^\ll_{\mu\nu}=0$, and 
$$
R^\ss_{~~\mu}-\fr{1}{2}\dd^\ss_\mu R=0~~,
$$
the standard equations for matterless gravity.  

\section{Linking Action}

We arrive at the action for the matrix, $M$, that links the two vierbein bundles
by treating it as the spin variable in a sigma model. This makes possible the gauging
away of local (orthochronous) Lorentz transformations in each bundle separately. The point is that
only the local distortion of one bundle relative to the other is of physical significance 
and not the particular choices of local frame in the two bundles. 

Because it is gauge invariant under local (orthochronous) Lorentz transformations in either bundle,
the action that achieves this is
\be
I_L=\fr{1}{16\pi F}\int d^4xJg^{\mu\nu}\Tr(j_\mu j_\nu)~~,
\ee
where $F$ is a new gravitational constant with the same dimensions as $G$~. The 
matrix valued current $j_\mu$ is given by
\be
j_\mu=(D_\mu M)M^{-1}~~,
\ee
More explicitly
\be
j_\mu^{ab}=(D_\mu M^a_{~~\brb})M^{\brb b}~~.
\ee

It is also convenient to define an alternative version of the current,
appropriate to the barred vierbein bundle, $\brj_\mu^{\bra\brb}$, as
\be
\brj_\mu^{\bra\brb}=\left(M^{-1}D_\mu M\right)^{\bra\brb}=M^{\bra}_{~~a}D_\mu M^{a\brb}~~.
\ee

We treat the vierbein, which enters through $g^{\mu\nu}$, the matrix, $M$, and
the connections $\om_{\mu ab}$ and $\bom_{\mu\bra\brb}$ as independent variables.
The result for the linking action from the vierbein variation is,
\be
\dd I_L=-\fr{1}{8\pi F}\int d^4xJ\dd e_{\ss c}
     \left(e^{c\nu}g^{\ss\mu}-\fr{1}{2}e^{c\ss}g^{\mu\nu}\right)\Tr(j_\mu j_\nu)~~,
\ee
From the left vierbein connection we have
\be
\dd I_L=\fr{1}{8\pi F}\int d^4xJ\dd\om_{\mu ab}j^{\mu ba}~~,
\ee
and from the right vierbein
\be
\dd I_L=-\fr{1}{8\pi F}\int d^4xJ\dd\bom_{\mu\bra\brb}\brj^{\mu\brb\bra}~~.
\ee
On varying the matrix $M$ we obtain
\be
\dd I_L=-\fr{1}{8\pi F}\int d^4xJ
   \Tr\left[(\dd MM^{-1})(D_\mu j^{\mu}-2C^\ll_{\ll\mu}j^{\mu})\right]~~.
\ee
The quantity $\dd MM^{-1}$ is an arbitrary element of the $SL(4,R)$ Lie algebra and is sufficiently 
general to identify the other factor in the trace which is also an element of the algebra.

\section{Matter Action}

We assume that matter is propagated in the vierbein background $\{\bre_{\mu\bra}\}$~.
This seems a consistent approach since it implies that matter behaves in a conventional way
in relation to the gravitational field it experiences. However the theory does
change the relationship of this observed gravitational field to the distribution
of matter density.

If we assume that the matter action does not depend explicitly on the right connection
$\bom_{\mu\bra\brb}$, then the most general form of the variation of the matter action
is
\be
\dd I_M=-\int d^4xJ\left(\dd e_{\ss c}T^{\ss c}+\Tr(\dd MM^{-1}U)\right)~~.
\ee
However because we assume that matter behaves conventionally in the effective gravitational
field we may write
\be
\dd I_M=-\fr{1}{2}\int d^4xJ\dd\brg_{\mu\nu}\BT^{\mu\nu}~~,
\ee
where $\BT^{\mu\nu}=\BT^{\nu\mu}$ is the symmetric energy momentum tensor for matter.
Since
\be
\brg_{\mu\nu}=\bre_{\mu\bra}\bre_\nu^{~~\bra}~~,
\ee
it follows that
\be
\dd I_M=-\int d^4xJ\dd\bre_{\mu\bra}\bre_\nu^{~~\bra}\BT^{\mu\nu}~~.
\ee
However since
\be
\dd\bre_{\mu\bra}=\dd e_{\mu a}M^a_{~~\bra}+e_{\mu a}\dd M^a_{~~\bra}~~,
\ee
it follows that
\be
T^{\ss c}=M^c_{~~\bra}\bre_\nu^{~~\bra}\BT^{\ss\nu}~~,
\ee
and hence
\be
T^\ss_{~~\ll}=e_{\ll c}T^{\ss c}=\bre_{\ll\bra}\bre_\nu^{~~\bra}\BT^{\ss\nu}=\BT^{\ss\nu}\brg_{\ll\nu}~~.
\ee
If we adopt the convention that barred quantities, that is those approriate to
the gravitational background of the matter, have spatial indices raised and lowered
with the barred metric we can define
\be
\BT^\ss_{~~\ll}=\BT^{\ss\nu}\brg_{\ll\nu}~~.
\ee
Hence we get the simple seeming result
\be
T^\ss_{~~\ll}=\BT^\ss_{~~\ll}~~.
\label{MTR1}
\ee
However it is important to recall that
\be
T^{\ss\tau}=T^\ss_{~~\ll}g^{\ll\tau}\ne\BT^{\ss\tau}~~.
\ee
In fact $T^{\ss\tau}$ is not necessecarily symmetric. This does not cause any difficultry.

We see also that
\be
U^{ba}=\left[e_\mu^{~~a}M^b_{~~\bra}\bre_\nu^{~~\bra}\BT^{\mu\nu}\right]~~,
\ee
where $\left[\cdots\right]$ indicates the projection of the contained quantity
onto the Lie algebra. Using the result $M^b_{~~\bra}=e^{b\ll}\bre_{\ll\bra}$
we can show that
\be
U^{ba}=e^{b\ll}e_\mu^{~~a}\BT^\mu_{~~\ll}-\fr{1}{4}\eta^{ba}\BT~~.
\label{MTR2}
\ee
where $\BT=\BT^\mu_{~~\mu}$~. Note that $T=T^\mu_{~~\mu}=\BT$~.

\section{Equations of Motion}

We obtain the equations of motion by requiring that the variation
of the total action is stationary. The result is
\be
\fr{1}{8\pi G}\left(R^\ss_{~~\rho}-\fr{1}{2}\dd^\ss_\rho R\right)
 -\fr{1}{8\pi F}\left(\Tr(j^\ss j_\rho)-\fr{1}{2}\dd^\ss_\rho\Tr(j^\ll j_\ll)\right)
                 -T^\ss_{~~\rho}=0~~.
\label{EQM1}
\ee
\be
\fr{1}{8\pi F}\left(D_\mu j^\mu-2C^\ll_{\ll\mu}j^\mu\right)+U=0~~.
\label{EQM2}
\ee
\be
j^{\ss[b,a]}=\fr{F}{G}e^{a\mu}e^{b\nu}
         \left(C^\ll_{\ll\mu}\dd^\ss_\nu-C^\ll_{\ll\nu}\dd^\ss_\mu+C^\ss_{\mu\nu}\right)~~.
\label{EQM3}
\ee
\be
\brj_\mu^{[\brb,\bra]}=0~~.
\label{EQM4}
\ee

Eq(\ref{EQM1}) implies that
\be
\fr{1}{8\pi G}R^\ss_{~~\rho}-\fr{1}{8\pi F}\Tr(j^\ss j_\rho)
                  =\left(T^\ss_{~~\rho}-\fr{1}{2}\dd^\ss_\rho T\right)~~.
\label{EQM1a}
\ee

\section{Bianchi Identity}

Just as in standard General Relativity it is necessary to check that the
theory satisfies the integrability conditions associated with the Bianchi identity.
In the presence of torsion these are changed to the following 
\be
R_{\ll\tau\mu\nu;\ss}+R_{\ll\tau\nu\ss;\mu}+R_{\ll\tau\ss\mu;\nu}=
 -2\left(R_{\ll\tau\rho\mu}C^\rho_{\nu\ss}+R_{\ll\tau\rho\nu}C^\rho_{\ss\mu}
+R_{\ll\tau\rho\ss}C^\rho_{\mu\nu}\right)~~,
\label{BI1}
\ee
where $;\mu$ indicates the covariant derivative $D_\mu$~. In the contracted version this is
\be
\left(R^\mu_{~~\ss}-\fr{1}{2}\dd^\mu_\ss R\right)_{;\mu}
=R^{\mu\nu}_{~~~\rho\ss}C^\rho_{\mu\nu}+2R^\mu_{~~\rho}C^\rho_{\ss\mu}~~.
\label{BI2}
\ee

If we set
\be
G^\ss_{~~\rho}=R^\ss_{~~\rho}-\fr{1}{2}\dd^\ss_\rho R~~,
\ee
and
\be
F^\ss_{~~\rho}=\Tr\left\{j^\ss j_\rho-\fr{1}{2}\dd^\ss_\rho j^\ll j_\ll\right\}~~,
\label{FDEF}
\ee
then eq(\ref{EQM1}) can be expressed as
\be
\fr{1}{8\pi G}G^\ss_{~~\rho}-\fr{1}{8\pi F}F^\ss_{~~\rho}-T^\ss_{~~\rho}=0~~.
\label{EQM1b}
\ee
We now take the covariant derivative, $D_\ss$, of this equation in order to test
the relationship of the Bianchi identity to the appropriate divergence law
for the matter energy-momentum tensor. Eq(\ref{BI2}) gives immediately $G^{\ss\ll}_{~~~;\ss}$~.
From eq(\ref{FDEF}) we obtain
\be
F^{\ss\ll}_{~~~;\ss}=\Tr(j^{\ss}_{~~;\ss}j^\ll)+g^{\ss\ll}\Tr(j^\mu (j_{\ss;\mu}-j_{\mu;\ss}))~~.
\ee
From eq(\ref{EQM2}) we have
\be
\Tr(j^\ss_{~~;\ss}j^\ll)=2C^\rho_{\rho\ss}\Tr(j^\ss j^\ll)-8\pi F\Tr(Uj^\ll)~~,
\ee
and using eq(\ref{EQM1a}) we find
\be
\Tr(j^\ss_{~~;\ss}j^\ll)=2\fr{F}{G}C^\rho_{\rho\ss}R^{\ss\ll}
    -16\pi FC^\rho_{\rho\ss}(T^{\ss\ll}-\fr{1}{2}g^{\ss\ll}T)-8\pi F\Tr(Uj^\ll)~~.
\ee
The cyclic property of traces allows us to write
\be
\Tr(j^\mu(j_{\ss;\mu}-j_{\mu;\ss}))=\Tr(j^\mu([D_\mu,D_\ss]M)M^{-1})~~.
\ee
If we introduce $\CR_{\mu\ss}$ and $\BCR_{\mu\ss}$ as Lie algebra matrices with the
definitions
\be
(\CR_{\mu\ss})_{ab}=R_{ab\mu\ss}~~~~~~~~\mbox{and}
                  ~~~~~~~~(\BCR_{\mu\ss})_{\bra\brb}=\BR_{\bra\brb\mu\ss}~~,
\ee
then eq(\ref{CURV2}) gives
\be
\Tr(j^\mu(j_{\ss;\mu}-j_{\mu;\ss}))
   =\Tr(j^\mu\CR_{\mu\ss}-\brj^\mu\BCR_{\mu\ss}-2C^\rho_{\mu\ss}j^\mu j_\rho)~~.
\ee
Eq(\ref{EQM4}) implies that the second term on the right yields zero under the trace.
From eq(\ref{EQM3}) and eq(\ref{EQM1a}) we obtain the result
\be
\Tr(j^\mu(j_{\ss;\mu}-j_{\mu;\ss}))
  =\fr{F}{G}\left(2C^\rho_{\rho\tau}R^\tau_{~~\ss}+2C^\rho_{\mu\ss}R^\mu_{~~\rho}
-C^\mu_{\tau\nu}R^{\tau\nu}_{~~~\mu\ss}\right)
+16\pi FC^\rho_{\mu\ss}(T^\mu_{~~\rho}-\fr{1}{2}\dd^\mu_\rho T)~~.
\ee 
Combining these results we find
\be
F^{\ss\ll}_{~~~;\ss}
=-\fr{F}{G}g^{\ss\ll}(2C^\rho_{\mu\ss}R^\mu_{~~\rho}
-C^\mu_{\tau\nu}R^{\tau\nu}_{~~~\mu\ss})-16\pi F(C^\rho_{\rho\ss}T^{\ss\ll}
-g^{\ss\ll}C^\rho_{\mu\ss}T^\mu_{~~\rho})
-8\pi F\Tr(Uj^\ll)~~.
\ee
The covariant divergence of eq(\ref{EQM1b}) then yields
\be
T^{\ss\ll}_{~~~;\ss}=\Tr(Uj^\ll)+2(C^\rho_{\rho\ss}T^{\ss\ll}
     -g^{\ss\ll}C^\rho_{\mu\ss}T^\mu_{~~\rho})~~,
\ee
alternatively
\be
T^\ss_{~~\ll;\ss}=\Tr(Uj_\ll)+2(C^\rho_{\rho\ss}T^\ss_{~~\ll}-C^\rho_{\mu\ll}T^\mu_{~~\rho})~~.
\ee

It is useful to note that
\be
j_{\ll ab}=-e_a^{~~\rho}(\GG^\tau_{\ll\rho}-\BGG^\tau_{\ll\rho})e_{\tau b}~~.
\ee
We have then
\be
\Tr(Uj_\ll)=-(\GG^\tau_{\ll\rho}-\BGG^\tau_{\ll\rho})(e_{\tau b}U^{ba}e_a^{~~\rho})~~.
\ee
Using eq(\ref{MTR2}) we find
\be
\Tr(Uj_\ll)=-(\GG^\tau_{\ll\rho}-\BGG^\tau_{\ll\rho})\BT^\rho_{~~\tau}~~,
\ee
where we have used the result $\Tr j_\ll=0$~. In turn this implies 
$\GG^\ss_{\tau\ss}=\BGG^\ss_{\tau\ss}$~. We conclude that
\be
D_\ss T^\ss_{~~\ll}=-(\GG^\tau_{\ll\rho}-\BGG^\tau_{\ll\rho})\BT^\rho_{~~\tau}
         +2(C^\rho_{\rho\ss}T^\ss_{~~\ll}-C^\rho_{\mu\ll}T^\mu_{~~\rho})~~.
\ee
The relationship between the covariant derivatives $D_\mu$ and $\BD_\mu$, implies
\be
D_\ss T^\ss_{~~\ll}=\BD_\ss T^\ss_{~~\ll}+(\GG^\ss_{\ss\tau}-\BGG^\ss_{\ss\tau})T^\tau_{~~\ll}
                       -(\GG^\tau_{\ss\ll}-\BGG^\tau_{\ss\ll})T^\ss_{~~\tau}~~.
\ee
We obtain then
\be
\BD_\ss T^\ss_{~~\ll}=(\BGG^\ss_{\ss\tau}-\GG^\ss_{\tau\ss})T^\tau_{~~\ll}
                           -(\BGG^\tau_{\ss\ll}-\BGG^\tau_{\ll\ss})T^\ss_{~~\tau}~~.
\ee
However we have $\GG^\ss_{\tau\ss}=\BGG^\ss_{\tau\ss}$, and find, replacing 
$T^\ss_{~~\ll}$ by $\BT^\ss_{~~\ll}$,
\be
\BD_\ss\BT^\ss_{~~\ll}=2\BC^\ss_{\ss\tau}\BT^\tau_{~~\ll}-2\BC^\tau_{\ss\ll}\BT^\ss_{~~\tau}~~.
\ee
It is readily checked that this is equivalent to the standard conservation law
\be
\hD_\ss\BT^\ss_{~~\ll}=0~~,
\ee
where $\hD_\ss$ is the covariant derivative incorporating the metric connection
for the metric $\brg_{\mu\nu}$~. We conclude therefore that the extended
equations we have proposed for VLC gravity are consistent with conservation of the
energy-momentum tensor for matter in the appropriate (barred) metric.

\section{Connection Structure}

One of the features of the theory is the r\^ole played by torsion.
Here we show that the equations of motion do permit an analysis
of the vierbein connections.

From eq(\ref{EQM4}) we can obtain $\bom_{\mu\bra\brb}$ in terms of the other variables.
We can therefore eliminate $\bom_{\mu\bra\brb}$ from the equations of motion.
We have
\be
j_{\mu ab}=M_a^{~~\bra}\brj_{\mu\bra\brb}M^\brb_{~~b}
                  =M_a^{~~\bra}\brj_{\mu\{\bra,\brb\}}M^\brb_{~~b}~~.
\ee
We find then
\be
j_{\mu ab}=\fr{1}{2}M_a^{~~\bra}\left(M_\bra^{~~c}\d_\mu M_{c\brb}
                    +M_\bra^{~~c}\om_{\mu c}^{~~~d}M_{d\brb}
                   +M_\brb^{~~c}\d_\mu M_{c\bra}
                    +M_\brb^{~~c}\om_{\mu c}^{~~~d}M_{d\bra}\right)M^\brb_{~~b}~~.
\label{CON1}
\ee
This equation exhibits $j_{\mu ab}$ as a linear function of $\om_{\mu ab}$~.
Eq(\ref{EQM3}) relates the torsion tensor $C^\ll_{\mu\nu}$ linearly to $j_{\mu ab}$~.
The resulting equation determines $\om_{\mu ab}$ in terms of $M^a_{~~\bra}$ and $e_{\mu a}$
and their derivatives. 

From the covariant constancy of $e_{\mu a}$ we have
\be
\GG^\ll_{\mu\nu}=e^{a\ll}\left(\d_\mu e_{\nu a}+\om_{\mu a}^{~~~b}e_{\nu b}\right)~~.
\ee
Hence
\be
C^\ll_{\mu\nu}=\fr{1}{2}e^{a\ll}\left(\d_\mu e_{\nu a}+\om_{\mu a}^{~~~b}e_{\nu b}
                                 -\d_\nu e_{\mu a}-\om_{\nu a}^{~~~b}e_{\mu b}\right)~~.
\ee
If we define
\be
C_{\ll\mu\nu}=g_{\ll\ss}C^\ss_{\mu\nu}~~,
\ee
and 
\be
\om_{\mu\ll\nu}=e_\ll^{~~a}e_\nu^{~~b}\om_{\mu ab}~~,
\ee
then we find
\be
\om_{\mu\ll\nu}=C_{\ll\mu\nu}-C_{\nu\mu\ll}+C_{\mu\ll\nu}+{\hat{\om}}_{\mu\ll\nu}~~,
\ee
where ${\hat{\om}}_{\mu\ll\nu}$ is the metric version of $\om_{\mu\ll\nu}$ and is given by
\be
{\hat{\om}}_{\mu\ll\nu}=\fr{1}{2}\left(e_\nu^{~~a}\d_\mu e_{\ll a}-e_\ll^{~~a}\d_\mu e_{\nu a}
                 +\d_\nu g_{\ll\mu}-\d_\ll g_{\nu\mu}\right)~~.
\ee

Eq(\ref{EQM3}) can be expressed in the form
\be
C_{\ss\mu\nu}+C^\ll_{\ll\mu}g_{\nu\ss}-C^\ll_{\ll\nu}g_{\mu\ss}=-X_{\ss\mu\nu}~~,
\ee
where
\be
X_{\ss\mu\nu}=\fr{G}{F}j_{\ss[\mu,\nu]}~~,
\ee
and
\be
j_{\ss\mu\nu}=e_\mu^{~~a}e_\nu^{~~b}j_{\ss ab}~~.
\ee
It follows that
\be
C_{\ss\mu\nu}=-X_{\ss\mu\nu}-\fr{1}{2}X^\ll_{\ll\mu}g_{\nu\ss}
                                   +\fr{1}{2}X^\ll_{\ll\nu}g_{\mu\ss}~~.
\ee
Finally we have solution
\be
\om_{\mu\ll\nu}=-X_{\ll\mu\nu}+X_{\nu\mu\ll}-X_{\mu\ll\nu}-X^\tau_{\tau\ll}g_{\mu\nu}
        +X^\tau_{\tau\nu}g_{\mu\ll}+{\hat{\om}}_{\mu\ll\nu}~~.
\label{CON2}
\ee

\section{Special Cases}
The above solution for $\om_{\mu\ll\nu}$ can be used rather straightforwardly 
in special cases. We consider a situation in which the vierbeins and the linking
matrix have diagonal forms, namely
\be
e_{\mu a}=A_a\eta_{\mu a}~~~~~~~~\mbox{and}~~~~~~~~M^a_{~~\bra}=\LL_a\dd^a_\bra~~.
\ee
It follows that
\be
e^{a\mu}=A^{-1}_a\eta^{a\mu}~~~~~~~~\mbox{and}~~~~~~~~M^{\bra}_{~~a}=\LL^{-1}_a\dd^\bra_a~~,
\ee
and that
\be
\bre_{\mu\bra}=\LL_aA_a\eta_{\mu\bra}~~~~~~~~\mbox{and}
               ~~~~~~~~\bre^{\bra\mu}=\LL^{-1}_aA^{-1}_a\eta^{\bra\mu}~~.
\ee
We enforce an obvious correspondence between the values of the symbols $\mu$, $a$ and $\bra$
to give meaning to the $\eta$ and $\dd$~ symbols. In an appropriate sense we can write
$A_a=A_{\bra}=A_\mu$ and $\LL_a=\LL_{\bra}=\LL_\mu$ where this is convenient.

In this special case we can compute $j_{\mu ab}$ from eq(\ref{CON1}) to yield
\be
j_{\ll ab}=\fr{\d_\ll\LL_a}{\LL_a}\eta_{ab}
                +\fr{1}{2}\om_{\ll ab}\left(1-\fr{\LL^2_a}{\LL^2_b}\right)
\ee
Hence
\be
j_{\ll[a,b]}=\fr{1}{4}\om_{\ll ab}\left(2-\fr{\LL^2_a}{\LL^2_b}-\fr{\LL^2_b}{\LL^2_a}\right)~~,
\ee
and
\be
X_{\ll\mu\nu}=\fr{G}{4F}\dd^a_\mu\dd^b_\nu A_aA_b\om_{\ll ab}
            \left(2-\fr{\LL^2_a}{\LL^2_b}-\fr{\LL^2_b}{\LL^2_a}\right)~~.
\label{CON3}
\ee
We also have
\be
{\hat{\om}}_{\mu ab}=\fr{\d_\nu A_a}{A_a}\left(\fr{A_a}{A_b}\right)\dd^\nu_b\eta_{\mu a}
                 -\fr{\d_\nu A_b}{A_b}\left(\fr{A_b}{A_a}\right)\dd^\nu_a\eta_{\mu b}~~.
\ee

\subsection{Flat Expanding Universe}
An application of the above equations is to a flat expanding universe where we
set
\be
A_0=1~~,~~~~A_i=A(t)~~,~~~~~~~~\mbox{and}~~~~~~~~\LL_0=\LL(t)~~,~~~~\LL_i=\LL_S(t)~~,
\ee
and $i$ runs over orthogonal spatial directions. The metric is therefore
\be
ds^2=dt^2-A^2(t)d\r^2~~.
\ee
The barred metric appropriate to matter is
\be
d\brs^2=\LL^2(t)dt^2-\LL^2_S(t)A^2(t)d\r^2~~.
\ee

If we adopt the convention that a coordinate label $\mu$ can take a time value 
which we denote by $t$ or acquire a spatial character which we denote by the
variables $x$, $y$ or $z$, then rotational invariance tells us that the only
non-vanishing components of $\om_{\ll\mu\nu}$ are $\om_{xty}=-\om_{xyt}$ and
that
\be
\om_{xty}=\om\dd_{xy}~~.
\ee  
The same is true for the metric connection, and we find that
\be
{\hat{\om}}_{xty}=A\dot{A}\dd_{xy}~~.
\ee

The same considerations of rotational invariance imply that only $X_{xty}=-X_{xyt}$
are nonvanishing and from eq(\ref{CON3}) we find
\be
X_{xty}=\fr{G}{4F}~\om_{xty}\left(2-\fr{\LL^2}{\LL_S^2}-\fr{\LL_S^2}{\LL^2}\right)~~.
\ee
We find also 
\be
X^\tau_{\tau t}=\fr{1}{A^2}\dd_{x'y'}X_{x'ty'}~~.
\ee
From eq(\ref{CON2}) we have then
\be
\om\dd_{xy}=\fr{G}{4F}\left(-2\om\dd_{xy}+\dd_{xy}A^2
\times\fr{1}{A^2}\dd_{x'y'}\om\dd_{x'y'}\right)
\left(2-\fr{\LL^2}{\LL_S^2}-\fr{\LL_S^2}{\LL^2}\right)+A{\dot A}\dd_{xy}~~.
\ee
Finally 
\be
\om=\fr{A{\dot A}}{E}~~,
\ee
where
\be
E=1-\fr{G}{4F}\left(2-\fr{\LL^2}{\LL_S^2}-\fr{\LL_S^2}{\LL^2}\right)~~.
\ee
We will use this result later when we consider flat FRW space-time.

\subsection{Static Rotationally Invariant Space-time} 

Another application of the formalism is to the equivalent of
the Schwarzschild solution. We will use isotropic coordinates,
so that
\be
A_0=A(r)~~,~~~~A_i=B(r)~~,~~~~~~~~\mbox{and}~~~~~~~~\LL_0=\LL(r)~~,~~~~\LL_i=\LL_S(r)~~.
\ee
The unbarred metric is then
\be
ds^2=A^2(r)dt^2-B^2(r)d\r^2~~,
\ee
and the barred metric is
\be
d\brs^2=\LL^2(r)A^2(r)dt^2-\LL_S^2(r)B^2(r)d\r^2~~.
\ee
Rotational invariance implies that only the components $\om_{xty}=-\om_{xyt}$, 
$\om_{ttx}=-\om_{txt}$ and $\om_{xyz}$ may be non-zero. In the case of the metric 
connection we find
\be
{\hat\om}_{xty}=0~~,
\ee
\be
{\hat\om}_{ttx}=AA'{\hat r}_x~~,
\ee
where ${\hat r}_x$ is the $x$-component of the radial unit vector.
We have also
\be
{\hat\om}_{xyz}=BB'\left({\hat r}_y\dd_{xz}-{\hat r}_z\dd_{xy}\right)~~.
\ee
The components of $X_{\ll\mu\nu}$ that may be non-zero have the values,
\be
X_{xty}=\fr{G}{4F}\om_{xty}\left(2-\fr{\LL^2}{\LL_S^2}-\fr{\LL_S^2}{\LL^2}\right)~~,
\ee
\be
X_{ttx}=\fr{G}{4F}\om_{ttx}\left(2-\fr{\LL^2}{\LL_S^2}-\fr{\LL_S^2}{\LL^2}\right)~~,
\ee
and
\be
X_{xyz}=0~~.
\ee
Solving for $\om_{\ll\mu\nu}$ we find a pattern similar to that for
the metric connection,
\be 
\om_{xty}=0~~,
\ee
\be
\om_{ttx}=\fr{AA'}{P}{\hat r}_x~~,
\ee
and
\be
\om_{xyz}= Q\left({\hat r}_y\dd_{xz}-{\hat r}_z\dd_{xy}\right)~~,
\ee
where
\be
P=1+\fr{G}{4F}\left(2-\fr{\LL^2}{\LL_S^2}-\fr{\LL_S^2}{\LL^2}\right)~~,
\ee
and
\be
Q=BB'+\fr{G}{4F}\left(2-\fr{\LL^2}{\LL_S^2}-\fr{\LL_S^2}{\LL^2}\right)\fr{B^2A'}{AP}~~.
\ee
We will use these results in analysing the Schwarzschild-type metric in VLC theory.

\section{Weak Field Limit}
The weak field limit has the form
\be
e_{\mu a}=e^{(0)}_{\mu a}+h_{\mu a}~~,
\ee
where
\be
e^{(0)}_{~~\mu a}e^{(0)~a}_{~~\nu}=\eta_{\mu\nu}~~,
\ee
Similarly we can set
\be
\bre_{\mu \bra}=\bre^{(0)}_{\mu \bra}+\brh_{\mu \bra}~~,
\ee
where
\be
\bre^{(0)}_{~~\mu \bra}\bre^{(0)~\bra}_{~~\nu}=\eta_{\mu\nu}~~.
\ee
The connections $\om_{\mu ab}$ and $\bom_{\mu\bra\brb}$ are first order quantities.
The matrix $M$ has the form
\be
M^a_{~~\bra}=M^{(0)a}_{~~~~~\bra}+m^a_{~~\bra}~~,
\ee
where $M^{(0)}$ is a Lorentz transformation. We have then
\be
M^{(0)~\bra}_{~~~~~~a}=M^{(0)~~\bra}_{~~~a}~~.
\ee
We can use $M$ and its inverse and $e$ and $\bre$ to convert superfixes and suffixes
between the various bases. For example we have
\be
m^a_{~~b}=m^a_{~~\bra}M^{(0)~\bra}_{~~~~~~b}~~.
\ee
The requirement that $\det M=1$ implies that
\be
m^a_{~~a}=m^\bra_{~~\bra}=m^\mu_{~~\mu}=0~~.
\ee
The relationship between $e$ and $\bre$ implies that
\be
\brh_{\mu\bra}=h_{\mu\bra}+m_{\mu\bra}~~,
\ee
together with corresponding equations in other bases.

To lowest order
\be
R_{ab\mu\nu}=\d_\mu\om_{\nu ab}-\d_\nu\om_{\mu ab}~~.
\ee
Hence
\be
R^{\ss}_{~~\mu}=e^{(0)a\ss}e^{(0)b\nu}R_{ab\mu\nu}~~,
\ee
so that
\be
R^{\ss}_{~~\mu}=\d_\mu\om_{\nu}^{~~\ss\nu}-\d_{\nu}\om_\mu^{~~\ss\nu}~~,
\ee
and
\be
R=2\d_\mu\om_\nu^{~~\mu\nu}~~.
\ee
Again in the lowest order approximation
\be
j_{\mu ab}=\d_\mu m_{ab}+\om_{\mu ab}-\bom_{\mu ab}~~.
\ee
If we convert to the coordinate basis we have
\be
j_{\mu\ll\tau}=\d_\mu m_{\ll\tau}+\om_{\mu\ll\tau}-\bom_{\mu\ll\tau}~~.
\ee
We can also evaluate $\brj_\mu$~. In this lowest approximation it
coincides with $j_\mu$~. From the equation of motion we have
\be
\brj_{\mu[\ll,\tau]}=j_{\mu[\ll,\tau]}=0~.
\ee
That is
\be
\d_\ss m_{[\ll,\tau]}+\om_{\ss\ll\tau}-\bom_{\ss\ll\tau}=0~~.
\label{CON4}
\ee
This result tells us that the torsion in the unbarred, gravitational vierbein
bundle is zero.

Eq(\ref{EQM2}) yields
\be
\eta^{\mu\ss}\d_\mu\left(\d_\ss m_{\ll\tau}+\om_{\ss\ll\tau}
                        -\bom_{\ss\ll\tau}\right)=-8\pi FU_{\ll\tau}~~,
\ee
where
\be
U_{\ll\tau}=\BT_{\ll\tau}-\fr{1}{4}\eta_{\ll\tau}\BT~~.
\ee
Making use of eq(\ref{CON4}) we obtain the result
\be
\eta^{\mu\ss}\d_\mu\d_\ss m_{\left\{\ll,\tau\right\}}
                 =-8\pi F\left(\BT_{\ll\tau}-\fr{1}{4}\eta_{\ll\tau}\BT\right)~~.
\ee
The remaining equations are to lowest order
\be
R_{\ss\ll}-\fr{1}{2}\eta_{\ss\ll}R=8\pi G\BT_{\ss\ll}~~,
\ee
and
\be
C^\ll_{\mu\nu}=\fr{1}{2}\left(\d_\mu h_\nu^{~~\ll}-\d_\nu h_\mu^{~~\ll}
        +\om_{\mu~~\nu}^{~\ll}-\om_{\nu~~\mu}^{~\ll}\right)=0~~.
\ee
In the present approximation $m_{[\mu,\nu]}$ can be absorbed by gauge
transformations of the form
\be
\om_{\ss\ll\tau}\rightarrow \om_{\ss\ll\tau}+\d_\ss\phi_{\ll\tau}~~,
~~~~~~~~\mbox{and}
~~~~~~~~\bom_{\ss\ll\tau}\rightarrow \bom_{\ss\ll\tau}+\d_\ss{\bar \phi}_{\ll\tau}~~.
\ee
We can assume therefore that in this approximation $m_{[\mu,\nu]}$ vanishes.
Therefore $m_{\mu\nu}$ may be assumed symmetric. It satisfies
\be
\d^2m_{\mu\nu}=-8\pi F(\BT_{\mu\nu}-\fr{1}{4}\eta_{\mu\nu}\BT)~~.
\label{WE1}
\ee
The gauge invariance referred to above means also the we are free to
choose $h_{\mu\nu}$ to be symmetric with the result that $\brh_{\mu\nu}$
is also symmetric.

Under these circumstances we can solve the vanishing torsion equation
to give
\be
\om_{\nu\ll\mu}=\d_\mu h_{\nu\ll}-\d_\ll h_{\nu\mu}~~.
\ee
Eq(\ref{EQM1}) now yields
\be
\d_\mu\d_\nu h^\nu_{~~\ss}+\d_\ss\d_\nu h^\nu_{~~\mu}-\d^2h_{\mu\ss}
-\d_\mu\d_\ss h^\nu_{~~\nu}-\eta_{\ss\mu}(\d_\nu\d_\tau h^{\nu\tau}-\d^2h^\nu_{~~\nu})
=8\pi G\BT_{\ss\mu}~~.
\ee
We now refine our coordinate system by choosing the harmonic gauge.
\be
g^{\mu\nu}\GG^\ll_{\mu\nu}=0~~.
\ee
In the lowest approximation it yields
\be
\d_\mu h^\mu_{~~\ll}=\fr{1}{2}\d_\ll h^\mu_{~~\mu}~~.
\label{GC}
\ee
The equation of motion then becomes
\be
-\d^2\left(h_{\mu\ss}-\fr{1}{2}\eta_{\mu\ss}h^\tau_{~~\tau}\right)=8\pi G\BT_{\mu\ss}~~,
\ee
or
\be
-\d^2 h_{\mu\ss}=8\pi G\left(\BT_{\mu\ss}-\fr{1}{2}\eta_{\mu\ss}\BT\right)~~.
\label{WE2}
\ee
\subsection{Static Case}

In a static situation
\be
\d^2=-\nabla^2~~,~~~~~~~~\mbox{and}~~~~~~~~\BT_{00}=\BT=\rho~~,
\ee
where $\rho$ is the density of matter.
We have then
\be
\nabla^2 h_{00}=4\pi G\rho~~,~~~~~~~~\mbox{and}~~~~~~~~~\nabla^2m_{00}=6\pi F\rho~~.
\ee
It follows that
\be
\nabla^2\brh_{00}=4\pi G_N\rho~~,
\ee
where 
\be
G_N=G+\fr{3}{2}F~~.
\ee
We interpret $\brh_{00}$ as the gravitational potential seen by matter.
Therefore $G_N$ can be identified with Newton's constant. Both 
gravitational constants $G$ and $F$ enter into the structure of weak gravity
but only in a particular linear combination.

\subsection{Gravitational Waves}

A time-dependent matter distribution can act as a source of gravitational
radiation. Clearly from eq(\ref{WE2}) we see that there are such waves of a conventional type.
However eq(\ref{WE1}) shows that we may also have unconventional gravitational radiation
associated with the oscillations of the field $m_{\mu\nu}$~. It is interesting to count 
the degrees of freedom introduced in this way.

Dealing first with the conventional degrees of freedom represented by $h_{\mu\nu}$
the standard argument is that the conservation of the energy momentum tensor implies that the
gauge condition eq(\ref{GC}) is maintained. This is a necessary consistency check.
Far from the source the wave, in the weak field limit, may be treated as a plane
wave of the form
\be
h_{\mu\nu}=A_{\mu\nu}e^{ik.x}~~,
\ee
and the gauge condition leads to the result
\be
k_\mu A^\mu_{~~\nu}-\fr{1}{2}k_\nu A^\tau_{~~\tau}=0~~.
\ee
These four conditions reduce the original ten degrees of freedom to six.
Finally four of these degrees of freedom may be removed by appropriate
coordinate transformations that preserve the gauge condition. This leaves the
standard two polarizations for the gravitational wave.

When we come to the degrees of freedom represented by the $m_{\mu\nu}$ we see that
the conservation law for the matter energy-momentum tensor implies
\be
\d^2\d_\mu m^\mu_{~~\nu}=2\pi F\d_\nu\BT~~.
\ee
However we know that
\be
\BT=\fr{1}{8\pi G}\d^2h^\tau_{~~\tau}~~.
\ee
Therefore
\be
\d^2\d_\mu m^\mu_{~~\nu}=\fr{F}{4G}\d^2\d_\nu h^\tau_{~~\tau}~~.
\ee
This implies, for waves that have their source in matter, that
\be
\d_\mu m^\mu_{~~\nu}-\fr{F}{4G}\d_\nu h^\tau_{~~\tau}=0~~.
\ee

If we assume we are far from the source so that
\be
m_{\mu\nu}=B_{\mu\nu}e^{ik.x}~~,
\ee
then we find
\be
k_\mu B^\mu_{~~\nu}-\fr{F}{4G}k_\nu A^\tau_{~~\tau}=0~~.
\ee
The symmetric matrix $m_{\mu\nu}$ has ten degrees of freedom. However
one is removed by the vanishing trace condition. Four others are removed by the 
conditions above, leaving five remaining degrees of freedom. This implies that
there are seven degrees of freedom left in the combined $\{h_{\mu\nu},m_{\mu\nu}\}$-system.
However the matter metric does not see one of these degrees of
freedom. We can easily deduce the result that
\be
\d_\mu\brh^\mu_{~~\nu}-\fr{1}{2}\left(1+\fr{F}{2G}\right)\d_\nu \brh^\tau_{~~\tau}=0~~.
\ee
These four equations reduce the original ten degrees of freedom in $\brh_{\mu\nu}$  
to six. Because the coefficient of the second term departs from $1/2$ when
$F\ne 0$, we cannot find a change of coordinates that removes further degrees
of freedom. Matter interacting with the observed gravitational waves will
see only six degrees of freedom. This is the conventional result for non-standard
gravity theories.

\section{FRW Universe}
A strong motivation for the VLC theory is the possible modification
that the theory may make to the evolution of the early universe. We
examine this issue in this section.

The equation of motion, eq(\ref{EQM2}), yields the result
\be
\d_\mu j^\mu+\GG^\mu_{\ll\mu}j^\ll+\om_\mu j^\mu-j^\mu\om_\mu=-8\pi FU~~.
\label{FRW1}
\ee
In the present case the only non-vanishing component of $\GG^\mu_{\ll\mu}$ is
\be
\GG^\mu_{t\mu}=\fr{\dot J}{J}=3\fr{\dot A}{A}~~.
\ee
In taking the $\{00\}$ component of eq(\ref{FRW1}) we find in the present case
\be
\d_t\left(\fr{\dot\LL}{\LL}\right)+3\fr{\dot A}{A}\fr{\dot\LL}{\LL}
  -\fr{3}{2}\left(\fr{\dot A}{AE}\right)^2\left(\fr{\LL^2}{\LL^2_S}-\fr{\LL^2_S}{\LL^2}\right)
           =-8\pi F\left(\BT^t_{~~t}-\fr{1}{4}\BT\right)=-6\pi F\left(\rho+p\right)~~.
\ee
The spatial components of the equation yield the same information.
We find also
\be
R^t_{~~t}=-\fr{3}{A}\d_t\left(\fr{\dot A}{E}\right)~~,
\ee
and
\be
\Tr\left(j^tj_t\right)=\fr{4}{3}\left(\fr{\dot\LL}{\LL}\right)^2~~,
\ee
yielding the equation of motion
\be
-\fr{3}{A}\d_t\left(\fr{\dot A}{E}\right)-\fr{G}{F}\fr{4}{3}\left(\fr{\dot\LL}{\LL}\right)^2
                          =4\pi G\left(\rho+3p\right)~~.
\ee
The spatial components are
\be
R^x_{~~y}=-\fr{1}{A}\d_t\left(\fr{\dot A}{E}\right)\dd_{xy}-2\left(\fr{\dot A}{AE}\right)^2\dd_{xy}~~,
\ee
and
\be
\Tr\left(j^xj_y\right)=-\fr{1}{2}\left(2-\fr{\LL^2}{\LL^2_S}-\fr{\LL^2_S}{\LL^2}\right)
                                   \left(\fr{\dot A}{AE}\right)^2\dd_{xy}~~,
\ee
leading to the equation of motion
\be
-\fr{1}{A}\d_t\left(\fr{\dot A}{E}\right)-2E\left(\fr{\dot A}{AE}\right)^2
                        =-4\pi G\left(\rho-p\right)~~.
\ee
These equations reduce to the standard equations for the flat FRW universe
when $F=0$~. They may be manipulated to yield the first order equation
\be
\left(\fr{\dot A}{E}\right)^2-\fr{2}{9}\fr{G}{F}\fr{A^2}{E}\left(\fr{\dot\LL}{\LL}\right)^2
                           =\fr{8\pi G}{3}\fr{A^2}{E}\rho~~.
\ee
It is also possible to verify, consistently with the Bianchi identities, 
that they imply the the appropriate energy conservation law
\be
\d_t\left(\fr{A^3}{\LL}\rho\right)+p\d_t\left(\fr{A^3}{\LL}\right)=0~~.
\label{FRW2}
\ee
If we set 
\be
c=\fr{\LL}{\LL_S}=\LL^{\fr{4}{3}}~~,
\ee
then c is the velocity, seen from the gravitational background, of a
signal travelling with the velocity of light in the matter background.
It is therefore the variable speed of light generated by the variable
light-cone structure of the theory.
We have
\be
\fr{\dot c}{c}=\fr{4}{3}\fr{\dot\LL}{\LL}~~.
\ee
The equations of motion can be rewritten in the form
\be
\d_t\left(\fr{\dot c}{c}\right)+3\fr{\dot A}{A}\fr{\dot c}{c}
-2\left(\fr{\dot A}{AE}\right)^2\left(c^2-\fr{1}{c^2}\right)=-8\pi F\left(\rho+p\right)~~,
\label{FRW3}
\ee
and
\be
E\left(\fr{\dot A}{AE}\right)^2-\fr{1}{8}\fr{G}{F}\left(\fr{\dot c}{c}\right)^2
                  =\fr{8\pi G}{3}\rho~~.
\label{FRW4}
\ee

\subsection{Steady Expansion}

An interesting consequence of these equations is that if the pressure
satisfies
\be
p=\ll\rho~~,
\ee
where $\ll$ is a constrant then there  exist solutions with constant values of $c\ne 1$~. We find
\be
\left(\fr{\dot A}{AE}\right)^2\left(c^2-\fr{1}{c^2}\right)=4\pi F(1+\ll)\rho~~,
\ee
and
\be
E\left(\fr{\dot A}{AE}\right)^2=\fr{8\pi G}{3}\rho~~.
\label{SE1}
\ee
We have then
\be
\fr{1}{E}\left(c^2-\fr{1}{c^2}\right)=\fr{3}{2}(1+\ll)\fr{F}{G}~~.
\ee
For relativistic matter $\ll=1/3$ and we obtain the quadratic equation
\be
\left(c^2\right)^2+2\left(1-\fr{2F}{G}\right)c^2-3=0~~,
\ee
with the physical solution
\be
c^2=c_0^2=-\left(1-\fr{2F}{G}\right)+2\sqrt{1-\fr{F}{G}+\fr{F^2}{G^2}}\simeq 1+\fr{F}{G}~~.
\ee
For pressureless matter $\ll=0$ and we obtain similarly
\be
c^2=c_0^2=-\fr{3}{5}\left(1-\fr{2F}{G}\right)
               +\fr{1}{2}\sqrt{\fr{36}{25}\left(1-\fr{2F}{G}\right)^2
                          +\fr{44}{5}}\simeq 1+\fr{3}{4}\fr{F}{G}~~.
\ee

We take the relativistic case as an illustration. From eq(\ref{SE1}) we see that the expansion
is controlled by
\be
\left(\fr{\dot A}{A}\right)^2=\fr{8\pi(GE_0)}{3}\rho~~,
\ee
where $E_0=E(c_0^2)$~. Because the speed of light is constant
we have from eq(\ref{FRW2})
\be
\rho\propto \fr{1}{A^4}~~.
\ee
This is exactly like the standard scenario for expansion in the early universe
with the change that $G_N$ has been replaced by $GE_0$~. It follows that
\be 
A\propto \sqrt{t}~~.
\ee
It is clearly true also that in the matter coordinates 
\be
\BA\propto \sqrt{\bar t}~~,
\ee
where $\BA=c_0^{-1/4}A$ and ${\bar t}=c_0^{3/4}t$~. This steady
expansion is essentially the same as the conventional case.

\subsection{Non-steady Expansion}

It very difficult to investigate the solutions of eqs(\ref{FRW3}) and (\ref{FRW4}) in general. However
some feeling for their nature can be obtained by examining 
solutions near the above steady expansion.
If we set
\be
c=c_0e^\xi~~,
\ee
and assume that $\xi$ is a small quantity and that we can neglect quantities
quadratic in $\xi$ and ${\dot \xi}$ so that eq(\ref{FRW4}) becomes 
\be
E\left(\fr{\dot A}{AE}\right)^2=\fr{8\pi G}{3}\rho~~.
\ee
If we examine the case of relativistic matter, then when we combine this result with
eq(\ref{FRW3}) we get
\be
{\ddot\xi}+3\fr{\dot A}{A}{\dot \xi}
 -2E\left(\fr{\dot A}{AE}\right)^2
    \left(\fr{1}{E}\left(c^2-\fr{1}{c^2}\right)-2\fr{F}{G}\right)=0~~.
\ee
Keeping only terms $O(\xi)$ we obtain
\be
{\ddot\xi}+3\fr{\dot A}{A}{\dot \xi}
 -2\left(\fr{\dot A}{AE_0}\right)^2
    \left(c_0^2+\fr{3}{c_0^2}\right)\xi=0~~,
\ee
where, in keeping with the approximation, $A$ may be assumed to be the 
the steady expansion solution obtained previously. As a further simplification
we will assume that $F$ is very small and set $F=0$ in the above equation.
We obtain then
\be
{\ddot\xi}+\fr{3}{2t}{\dot \xi}-\fr{2}{t^2}\xi=0~~.
\ee
This homogeneous equation has two linearly independent solutions
\be
\xi=t^\aa~~,~~~~~~~~\mbox{and}~~~~~~~~~\xi=\fr{1}{t^\bb}~~.
\ee
where $\aa=(\sqrt{33}-1)/4=1.1861$ and $\bb=(\sqrt{33}+1)/4$=1.6861~.
Using these solutions we can construct a range of scenarios for a (small) departure
from the steady expansion. A solution of the form 
\be
\xi=\fr{b}{t^\bb}~~,
\ee
will yield 
\be
c=c_0e^{b/t^\bb}~~,
\ee
which for $b>0$ results in the speed of light increasing in the far past and
tending to the steady expansion in the future. If $b<0$ then the speed of
light will increase to the steady value from a smaller value in the past.
If we choose a solution
\be
\xi=at^\aa
\ee
then when $a>0$ the speed of light will increase from the steady value
or when $b<0$ it will decrease from this value. By choosing arbitrary combinations
of the two solutions any mixture of these scenarios can be obtained. Of course
the analysis being perturbative, the results cannot be trusted beyond the point
where $\xi\simeq 1$, so the ultimate behaviour at very small or large times
remains unresolved.

\section{Static Rotationally Invariant Case}

We derive the equations for the static rotationally invariant system in
isotropic coordinates.  They can be used to study the equivalent of the 
Schwarzschild solution.  However we will confine our attention here to the 
asymptotic properties of the solution as a way of investigating the post-Newtonian 
approximation.

The equation of motion, eq(\ref{EQM2}), in the absence of matter yields
\be
\d_\mu j^\mu+\GG^\mu_{\ll\mu}j^\ll+\om_\mu j^\mu-j^\mu\om_\mu=0~~.
\ee
Making use of the fact that
\be
\GG^\mu_{\ll\mu}=\fr{\d_\mu J}{J}=\fr{\d_\mu A}{A}+3\fr{\d_\mu B}{B}~~,
\ee
we we see that
\be \GG^\mu_{t\mu}=0~~,~~~~~~~~\mbox{and}~~~~~~~~~\GG^\mu_{x\mu}=
                  \left(\fr{A'}{A}+3\fr{B'}{B}\right)\hr_x~~.
\ee
From the $\{00\}$ component of eq() we find
\be
\left(\fr{\LL'}{\LL}\right)'+\left(\fr{2}{r}+\fr{A'}{A}+3\fr{B'}{B}\right)\fr{\LL'}{\LL}
 -\fr{1}{2}\left(\fr{A'}{AP}\right)^2\left(\fr{\LL^2}{\LL_S^2}-\fr{\LL_s^2}{\LL^2}\right)=0~~.
\ee
The other non-trivial components yield the same equation. Using the definition
$F_{\mu\nu}=\Tr\{j_\mu j_\nu\}$ we find
\be
F_{tt}=\fr{A}{B}\left(\fr{A'}{P}\right)^2\left(\fr{\LL^2}{\LL_S^2}-\fr{\LL_S^2}{\LL^2}\right)~~,
\ee
and
\be
F_{xy}=\fr{4}{3}\left(\fr{\LL'}{\LL}\right)^2\hr_x\hr_y~~.
\ee
We also obtain for the relevant components of $R_{\mu\nu}$ the results
\be
R_{tt}=\fr{A}{B}\left[\left(\fr{A'}{BP}\right)'+\fr{2}{r}\left(\fr{A'}{BP}\right)
          -2\left(\fr{A'}{BP}\right)\left(\fr{Q}{B^2}\right)\right]~~,
\ee
and
\be
R_{xy}=R^{(1)}_{xy}+R^{(2)}_{xy}+R^{(3)}_{xy}~~,
\ee
where
\be
R^{(1)}_{xy}=-\left[\fr{B}{A}\left(\fr{A'}{BP}\right)'
                       +\left(\fr{Q}{B^2}\right)'\right]\hr_x\hr_y~~,
\ee
\be
R^{(2)}_{xy}=-\left[\left(\fr{Q}{B^2}\right)'+\fr{2}{r}\fr{Q}{B^2}\right]\dd_{xy}~~,
\ee
and
\be
R^{(3)}_{xy}=\left[\left(\fr{A'}{AP}\right)\fr{Q}{B^2}
             -\fr{1}{r}\fr{B}{A}\left(\fr{A'}{BP}\right)
              -\fr{1}{r}\left(\fr{Q}{B^2}\right)
         -\left(\fr{Q}{B^2}\right)^2\right]\left(\dd_{xy}-\hr_x\hr_y\right)~~.
\ee
From the $\{tt\}$ component of the equation of motion we obtain
\be
\left(\fr{A'}{BP}\right)'+\fr{2}{r}\left(\fr{A'}{BP}\right)
       -2\left(\fr{A'}{BP}\right)\left(\fr{Q}{B^2}\right)
               -\fr{G}{F}\left(\fr{A'}{P}\right)^2
              \left(2-\fr{\LL^2}{\LL_S^2}-\fr{\LL_s^2}{\LL^2}\right)=0~~.
\ee
If we contract the $\{xy\}$ components of the equation of motion with $\hr_y$ we obtain
\be
\fr{B}{A}\left(\fr{A'}{BP}\right)'+2\left(\fr{Q}{B^2}\right)'+\fr{2}{r}\left(\fr{Q}{B^2}\right)
      +\fr{4}{3}\fr{G}{F}\left(\fr{\LL'}{\LL}\right)^2=0~~.
\ee
On contracting with a unit vector orthogonal to $\hr_y$ we obtain
\be
\left(\fr{Q}{B^2}\right)'+\fr{3}{r}\left(\fr{Q}{B^2}\right)
                 +\fr{1}{r}\fr{B}{A}\left(\fr{A'}{BP}\right)
                   -\left(\fr{A'}{BP}\right)\fr{Q}{B^2}+\left(\fr{Q}{B^2}\right)^2=0~~.
\ee

\subsection{Asymptotic behaviour}

We assume that, at large $r$, the leading asymptotic behaviour of the solution
to these equations has the form
\be
\LL=1+\fr{\LL_1}{r}+\fr{\LL_2}{r^2}+O(\fr{1}{r^3})~~,
\ee
\be
A=1+\fr{A_1}{r}+\fr{A_2}{r^2}+O(\fr{1}{r^3})~~,
\ee
and
\be
B=1+\fr{B_1}{r}+\fr{B_2}{r^2}+O(\fr{1}{r^3})~~.
\ee
On evaluating the two leading order contributions to each of the equations 
we find the results
\be
A_1+B_1=0~~.
\ee
This result holds in the conventional theory.
\be
\LL_2=\fr{1}{2}\LL_1^2+A_1\LL_1~~,
\ee
\be
A_2=\fr{1}{2}A_1^2~~,
\ee
\be
B_2=-\fr{1}{4}A_1^2-\fr{1}{6}\fr{G}{F}\LL_1^2~~.
\ee

If we compare this asymptotic behaviour with the weak field approximation
then we see that 
\be
A_1=-GM~~,~~~~~~~~~\mbox{ and}~~~~~~~~~ \LL_1=-\fr{3}{2}FM~~,
\ee
where $M$ is the mass of the object at the centre of the spherically symmetric system.
The implication of these results for the barred (matter) metric is easily checked.
We find
\be
A^2\LL^2\simeq 1-\fr{2G_N}{r}+\fr{(2G_N^2+3GF)M^2}{r^2}~~,
\ee
and
\be
B^2\LL_S^2\simeq 1+\fr{2(G_N-F)M}{r}~~.
\ee
The standard post-Newtonian parametrisation is
\be
A^2\LL^2\simeq 1-\fr{2G_N}{r}+\fr{2\bb G_N^2M^2}{r^2}~~,
\ee
and
\be
B^2\LL_S^2\simeq 1+\fr{2\gg G_NM}{r}~~,
\ee
where for standard General Relativity we have $\bb=\gg=1$~. In the theory we have
outlined we find
\be
\bb=1+\fr{3}{2}\fr{GF}{G_N^2}\simeq 1+\fr{3}{2}\fr{F}{G_N}+O(F^2)~~,
\ee
and
\be
\gg=1-\fr{F}{G_N}~~.
\ee
The time-delay results yield $\gg=1.00\pm .002$ \cite{TD,CW1} which places a strong restriction
on $f/G_N$. An even stronger restriction results from 
the very accurate VLBI data we have $\gg=1.000\pm.00032$ \cite{VLBI,CW2}.
It follows immediately that $F<2\times 10^{-3}G_N$ with the consequence 
$\bb-1< 3\times 10^{-3}$~. The appropriate conclusion is that on the scale
of $G$ or $G_N$, $F$ is very small. This however does not prevent it having an effect
in the very early universe as indicated in the discussion of the FRW model.
Nevertheless these experiments could be taken to rule out the model as we 
have presented it here. Variations of the model in which extra terms are introduced
into the linking Lagrangian that give a mass to the new degrees of freedom
will lead to them having a finite range. This would leave the long distance asymptotic 
behaviour of the model coincident with that of standard General Relativity
at least in the weak field limit.
It would still permit modifications of the standard theory at short distances.
In particular it would allow the existence of very transparent modifications
of the effective metric which could be interpreted as a novel form of matter. 
The potential of such a model to explain the presence of dark matter in the
universe is intriguing and requires further study.

\section{Stability}

Because of the indefinite character of the local Lorentz metric it is not obvious that
the degrees of freedom in the linking field $M$ all contribute positively 
to the energy, in other words that they correspond to a stable theory. 
While we cannot at present give a final answer to this question we find it illuminating
to consider the issue in the case of two dimensions. Here the matrix $M$
has the form
\be
M=\left(\begin{array}{cc}a&b\\c&d\end{array}\right)~~,
\ee
with $ad-bc=1$~. The quantity $f=\Tr MM^T$ is invariant under separate left
and right Lorentz transformations. We have
\be
f=a^2+d^2-b^2-c^2~~.
\ee
If we set 
\be
M=LSL'~~,
\ee
where $L$ and $L'$ are independent Lorentz transformations then of course
\be
f=\Tr MM^T=\Tr SS^T~~.
\ee
If we choose the special form for $S$
\be
S=\left(\begin{array}{cc}\ll&0\\0&\ll^{-1}\end{array}\right)~~,
\ee
then
\be
f=\ll^2+\ll^{-2}\ge 2~~.
\ee
If we choose
\be
S=\left(\begin{array}{cc}\cos\theta&-\sin\theta\\\sin\theta&\cos\theta\end{array}\right)~~,
\ee
then
\be
f=2\cos^2\theta-2\sin^2\theta=2\cos 2\theta~~.
\ee
In this case
\be
-2\le f\le 2~~.
\ee
 third possibility is
\be
S=\left(\begin{array}{cc}0&-\ll^{-1}\\ \ll&0\end{array}\right)~~.
\ee
We then find
\be
f=-\left(\ll^2+\ll^{-2}\right) \le -2~~.
\ee
The three ranges together cover all possibilities for the value of $f$~.
We therefore can conclude that the matrix $M$
will be gauge equivalent to the appropriate special form $S$ provided that
form is chosen according to the value of $f$~. Indeed
it is possible to check explicitly that given $f$ the real matrices $L$, $L'$
and $S$ can be computed provided $S$ is chosen in the appropriate manner.

It is therefore clear that the object that results from equivalencing
matrices $M\in SL(2,R)$ by simultneous left and right Lorentz transformations,
although made up of pieces of manifold, is not itself a manifold. With
the sign conventions we have adopted for the linking action, oscillations
for which the field strengths maintain $f\ge 2$ will yield positive energies. 
The remaining two cases yield negative and then positive energy oscillations.

Similar yet more complicated possibilities exist for the four dimensional
case. The completely positive mode case corresponds to matrices $M$
that are Lorentz gauge equivalent to the special form $S$
where
\be
S=\left(\begin{array}{cccc}\LL_0&0&0&0\\0&\LL_1&0&0\\0&0&\LL_2\\0&0&0&\LL_3\end{array}\right)~~,
\ee
where $\LL_0\LL_1\LL_2\LL_3=1$~. This was the assumption we made implicitly
in the special cases we worked out above. It is not known whether solutions 
of the equations of motion remain in the purely in the positive sector.
It is not known whether it is important for this to be the case. In the sense we have
elucidated here, therefore, the stability of the theory remains undecided.
However the examples discussed in detail show that possible solutions with stable
characteristics exist.

\section{Conclusions}

We have shown that it is possible to formulate a purely geometrical theory with a variable
speed of light in a covariant fashion. The theory gives a meaning to
variations of the speed of light by identifying two vierbein bundles
one associated with gravity and one associated with matter. Matter
moves according a conventional Lagrangian in its own gravitational back ground.
Its energy momentum tensor is conserved in the usual way. In this way it similar to
other bi-metric theories \cite{CM,R1,R2,R3}
Gravity is also formulated in a conventional way in terms of its own 
vierbein bundle. The extra dynamics is supplied by a linking action that
treats the matrix relating the two bundles as a dynamical variable.
The form of the action is based on a sigma-model construction and requires
the introduction of a new coupling constant with the same dimensions as the
Newtonian constant. The action is not unique and can be modified by the 
addition of terms that effectively give a mass to the new degrees of freedom
at the expense of introducing new parameters. 

In this paper we have examined the the minimal version of the theory
without extra mass inducing terms. As a result we find the asymptotic
properties of the theory are modified relative to the standard Einstein theory.
The parameter $\gg$ of the post-Newtonian approximation is not equal to unity.
This is inconsistent with the time-delay and VLBI measurements \cite{TD,CW1,VLBI,CW2}. 
Since our theory 
reduces to the standard one when the new coupling constant, $F$, is set to zero we 
can of course achieve agreement with experiment by reducing this coupling to 
a sufficiently low value. A very low value for $F$ does not by itself preclude
strong effects in the very early universe. However the most natural conclusion 
is that the a departure from standard GR along the lines suggested is unlikely.
It is possible that with the introduction of mass terms
in the linking action the resulting modification of the asymptotic properties 
will leave the theory with essentially standard properties at long distances and
with modifications only at short distances. We have not studied this point
in this paper.

We have also shown that a (flat) FRW universe can be found in which it
is possible to see the speed of light behaving in the way envisaged by
previous authors. This would indeed permit the uniformisation of
cosmic temperature associated with the cosmic microwave background. 
However we have not been able to analyse the model sufficiently
closely to establish precisely what may happen in the very early, or
indeed late, universe. The results must therefore be treated as suggestive
rather than definitive. Nevertheless the compatibility of the variable
speed of light idea with covariance and energy and momentum conservation for
matter has been established within a completely geometrical theory.

Finally we must draw attention to the fact that the stability of the theory
against the production of negative energy densities associated with the new
degrees of freedom has not been satisfactorily established. It would be interesting 
to check that matter oscillations do not give rise to negative energy radiation.

\section*{Acknowledgements}
I thank Jonathon Evans for helpful and clarifying discussions.
\pagebreak


\begin{thebibliography}{99}
\bibitem{AM} A. Albrecht and J. Magueijo, Phys. Rev. D {\bf 59}, 043516, 1999
\bibitem{B1} J. D. Barrow, Phys. Rev. D {\bf 59}, 043515, 1999
\bibitem{BM1} J. D. Barrow and J. Magueijo, Phys. Letts. B {\bf 447}, 246-250, 1999
\bibitem{BM2} J. D. Barrow and J. Magueijo, Class. Quantum Grav. {\bf 16}, 1435-1454, 1999
\bibitem{BM3} J. D. Barrow and J. Magueijo, astro-ph/9907354
\bibitem{B2} J. D. Barrow, astro-ph/9811022
\bibitem{M1} J. W. Moffat, International Journal of Modern Physics D {\bf 2}, 351-365, 1999
\bibitem{M2} J. W. Moffat, astro-ph/9811390
\bibitem{CM} M. A. Clayton and J W Moffat, astro-ph/9812481
\bibitem{R1} N. Rosen, J. Gen. Rel. and Grav. {\bf 4}, 435-447, 1974
\bibitem{R2} N. Rosen, Ann. Phys. (N.Y.) {\bf 84}, 455-473, 1977
\bibitem{R3} N. Rosen, J. Gen. Rel. and Grav. {\bf 9}, 339-351, 1979
\bibitem{ITD} I T Drummond and S J Hathrell, Phys. Rev. D {\bf22}, 343, 1980
\bibitem{KD1} A. D. Dolgov and I. B. Kriplovich (Novosibirsk, IYF). IYF-81-53, 1981
\bibitem{KD2} A. D. Dolgov and I. B. Kriplovich (Novosibirsk, IYF). IYF 83-53, 1983
\bibitem{SD1} R. D. Daniels and G. M. Shore, Phys. Lett. B {\bf 367}, 75, 1996
\bibitem{SD2} R. D. Daniels and G. M. Shore, Nucl. Phys. B {\bf 425}, 634, 1994
\bibitem{LT} J. I. Latorre, P. Pascual and R. Tarrach, Nucl. Phys. B {\bf 437}, 1995
\bibitem{K} I. B. Kriplovich, Phys. Lett. B {\bf 346}, 251, 1995
\bibitem{GS} G. M. Shore, Nucl. Phys. B {\bf 460}, 379, 1996
\bibitem{TD} R, D. Reasenberg, Phil.Trans. Roy. Soc. {\bf 310}, 227, 1983
\bibitem{CW1} C. M. Will, ``Theory and experiment in gravitational physics'', 
Cambridge University Press, 1993
\bibitem{VLBI} T. M. Eubanks {\it et al}, Bull. Am. Phys. Soc. Abstract \# K 11.05 (1997)
\bibitem{CW2} C. M. Will, ``Confrontation between General Relativity and
Experiment: a 1998 update'', gr-qc/9811036
\end{thebibliography}
\end{document}